%% file: ms.tex
\documentclass{article}
\usepackage{spconf, graphicx}
\usepackage{subfiles}
\usepackage[fleqn]{amsmath}
\usepackage{hyperref} 
\usepackage{enumitem}
\usepackage{url}
\usepackage{amsfonts}
\usepackage{pdfpages}

\title{Self-Supervised Annotation of Seismic Images using Latent Space Factorization}
\twoauthors{Oluwaseun Joseph Aribido, Ghassan AlRegib
                      }
                            {School of Electrical and Computer Engineering\\
                            Georgia Institute of Technology,\\
                            Atlanta, GA,  30332-0250, USA \\
                            \{oja, alregib\}@gatech.edu}
                            {Mohamed Deriche}
                            {Department of Electrical Engineering\\
                            King Fahd University of \\
                            Petroleum and Minerals\\
                            Dhahran, Saudi Arabia\\
                            mderiche@kfupm.edu.sa \\}

\begin{document}
%
\begin{titlepage}

\onecolumn 


\begin{description}[labelindent=-1cm, leftmargin=2cm, style=multiline]

\item[\textbf{Citation}]{O. J. Aribido, G. AlRegib, and M. Deriche, ``Self-Supervised Annotation of Seismic Images Using Latent Space Factorization," in IEEE International Conference on Image Processing (ICIP), Abu Dhabi, United Arab Emirates, Oct. 2020.}


\item[\textbf{Review}]{Date of publication: 25 October 2020}

\item[\textbf{Data and Codes}]{\url{https://github.com/olivesgatech/Latent-Factorization}}

\item[\textbf{Bib}] {@inproceedings\{Aribido2020Self,\\ 
author=\{Aribido, Oluwaseun and AlRegib, Ghassan and Deriche, Mohamed\},\\ 
booktitle=\{2020 IEEE International Conference on Image Processing (ICIP)\},\\ 
year=\{2020\},\}
} 


\item[\textbf{Copyright}]{\textcopyright 2020 IEEE. Personal use of this material is permitted. Permission from IEEE must be obtained for all other uses, in any current or future media, including reprinting/republishing this material for advertising or promotional purposes,
creating new collective works, for resale or redistribution to servers or lists, or reuse of any copyrighted component
of this work in other works. }

\item[\textbf{Contact}]{\href{mailto:oja@gatech.edu}{oja@gatech.edu}  OR \href{mailto:alregib@gatech.edu}{alregib@gatech.edu}\\ \url{https://ghassanalregib.info} \\ }
\end{description}

\thispagestyle{empty}
\newpage
\clearpage
\setcounter{page}{1}

\twocolumn

\end{titlepage}

\maketitle
\begin{abstract}
Annotating seismic data is expensive, laborious and subjective due to the number of years required for seismic interpreters to attain proficiency in interpretation. In this paper, we develop a framework to automate annotating pixels of a seismic image to delineate geological structural elements given image-level labels assigned to each image. Our framework factorizes the latent space of a deep encoder-decoder network by projecting the latent space to learned sub-spaces. Using constraints in the pixel space, the seismic image is further factorized  to reveal confidence values on pixels associated with the geological element of interest. Details of the annotated image are provided for analysis and qualitative comparison is made with similar frameworks.
\end{abstract}

\subfile{introduction}
\subfile{proposed_model}

\subfile{adversarial_training}
\subfile{results_and_discussion}



\end{document}

%% file: introduction.tex
\section{Introduction}
\label{sec:intro}                                      

In recent times, self-supervised learning research aims at teaching models to learn intrinsic information from dataset with as few labels as possible. Particularly, in image segmentation tasks, research efforts are geared towards training models to `understand' the data and to generalize well on out-of-distribution data by `paying attention' to correlations and distributions of deep intrinsic properties of the data. In parallel, similar research on unsupervised learning of seismic facies are well explored. In the seismic literature, attributes are extracted from the seismic volume to learn the physics of the seismic reflections \cite{di2019reflector, di2019three, shafiq2017texture, shafiq2018role,alaudah2016generalized}. Afterwards, an unsupervised learning algorithm - usually K-means and Konohen's self Organizing maps are used to cluster relevant attributes together \cite{barnes2002investigation, de2007unsupervised}. However,  limited contribution has been made to self-learning geological structures or facies in deep learning frameworks. 
For instance, Alaudah et. al, \cite{alaudah2018structure} developed a weakly-labelled approach using Non-negative matrix factorization to find the best features that characterize image classes. Initially, a texture similarity image retrieval measure was used to extract 99$\times$99 image data from the F3 block dataset. Four classes were extracted: Horizons, Faults, Chaotic and Salt Domes. These images were stretched into vectors and factorized using a Non-Negative Matrix Factorization (NMF) algorithm, to extract important features that separate them into individual classes. The authors then setup an optimization problem to find the best features that discriminate between the image classes. These features are then shown to delineate geological structures within the images. However, this was not a deep learning based framework. 
Similarly, Shafiq et. al, \cite{shafiq2018towards} trained a sparse auto-encoder on a natural-image dataset and showed that the auto-encoder learns features that are similar to fine edges observed in seismic reflection images. These learned weights are then used to delineate the edges of a salt-dome, analogous to transfer learning in computer vision tasks. The property of learning weights that could be further tuned was demonstrated on the latent space of a 3-D encoder-decoder by \cite{dubrovina2019composite} in which the latent space was projected to various subspaces and each subspace mapped to various parts of an object. Our framework defers from  \cite{dubrovina2019composite} considerably in that in \cite{dubrovina2019composite} the projected subspaces were learned from strongly labelled parts. In our framework, the model figures out the annotated parts itself using imposed constraints and learns the relevant subspace simultaneously. In addition, we do not use a spatial transformation network in the model pipeline. The factorized parts of our image are managed carefully using training techniques discussed in the proposed model.  

The summary of our contribution in this paper are as follows: first, we use two projection matrices to learn two orthogonal subspaces of the latent space of our encoder-decoder model. Then we map these learned subspaces to geological elements in the pixel space of our input images. The orthogonal subspaces were learned using self-supervision. Secondly, although our dataset source had image level labels provided, we learn important geological models without using those labels (weak or strong) in our framework which makes our problem setup self-supervised.

%% file: proposed_model.tex
\section{Proposed Model}
\label{sec:proposed_model}

The dataset used in this paper is based off the four classes extracted in \cite{alaudah2018structure}. The source of the dataset is the F3 block volume obtained from the Offshore North Sea of the Netherlands and published by dGB Earth Sciences Inc. The dataset was pre-processed in \cite{alaudah2018structure} and applied to our model directly. First, few exemplar images of each class is selected and a similarity based technique is used to retrieve similar images. The similarity based technique is based on curvelet transforms introduced by \cite{starck2002curvelet}. Curvelet transforms are frequency-based fine strokes that capture fine-edges in the images while ignoring random noise efficiently. These coefficients are multi-scaled fine-strokes and they are particularly relevant in image de-noising problems. In textural images comparison, \cite{alfarraj2016content} demonstrated that curvlet coefficients achieve state-of-the-art metrics in differentiating texture images. Howbeit, post-migrated seismic images bear strong similarity to texture images and this was evidently demonstrated by \cite{long2015characterization}. As such, \cite{alaudah2018structure} selected 500 images for each class to make a total of $2000$ training images. These $2000$ images are part of $17000$ images in the Landmass-1\footnote{https://ieee-dataport.org/open-access/landmass} dataset.

\subsection{Our Deep Learning Framework}
\label{model}
To learn pixel-level annotation that identifies a geologic element of interest, we propose a composite architecture. First, we assume each image contains a dominant geologic structure and the rest of the image is background. It is important to note that not all $2000$ images are nicely cutout to feature only one geologic structure, but we make this safe assumption due to the label class assigned by our image retrieval technique. However, this labelled class could also be wrong in a few instances since its extraction is weakly-supervised.

\begin{figure}
       \centering
       \includegraphics[width=0.45\textwidth]{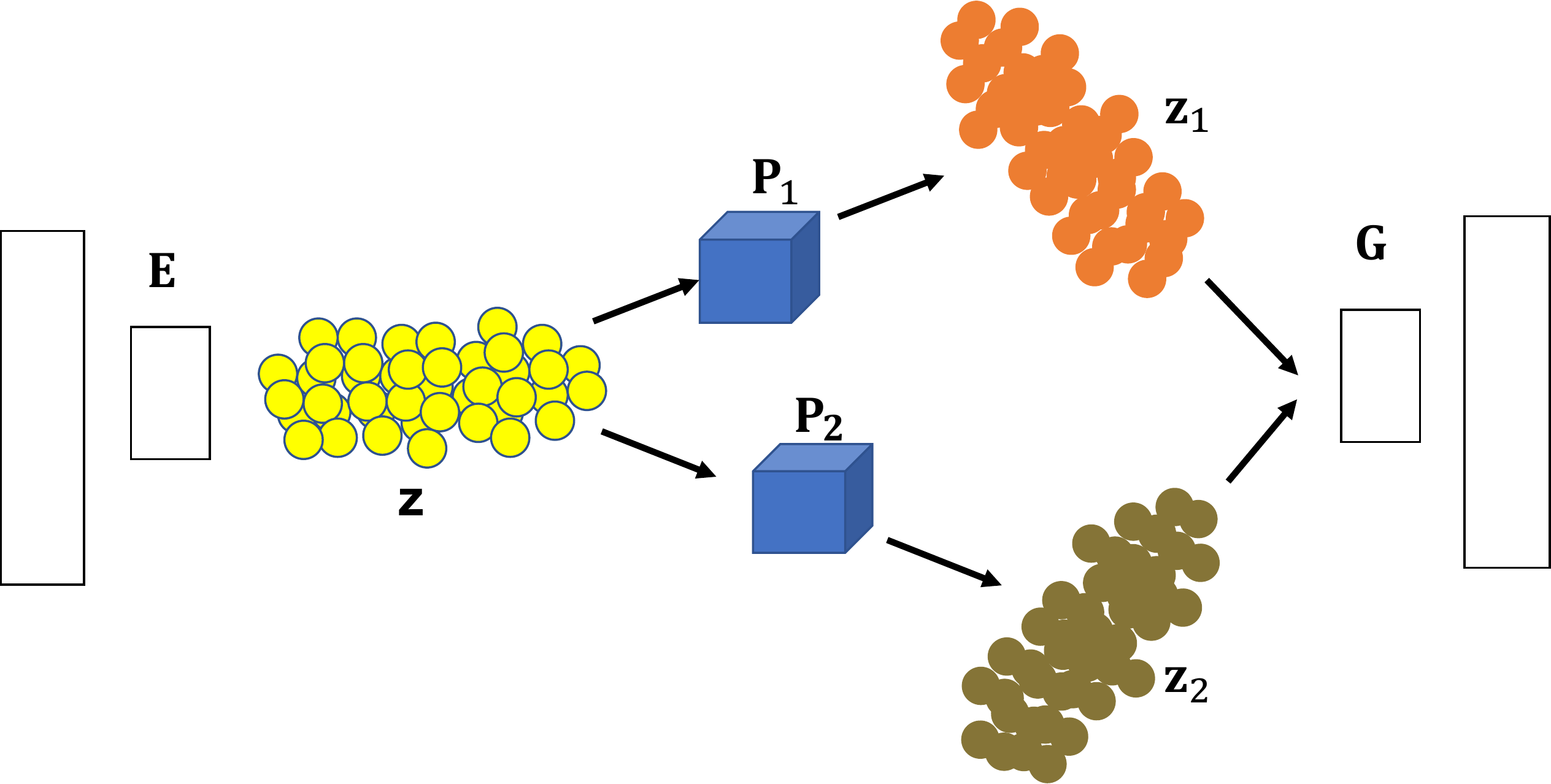}
       \caption{Latent Space Projection using Matrices: $\mathcal{P}_1$ and $\mathcal{P}_2$}
       \label{fig:latent_space}
   \end{figure} 

The model comprises a 5-layer encoder $(\mathbf{E})$ coupled with a 5-layer decoder $(\mathbf{G})$. Each layer is a 2D-convolutional layer with a kernel size of $5\times5$ and a stride of 2 followed by a batch-norm layer. No MaxPool or UnPooling layers are used, neither are there any fully connected layers. Between the encoder and decoder, there is a bottle neck denoted by $ \mathbf{z}$, a $1024 \times 1$ latent vector; which is $\mathbf{E}$'s output vector. During training, we pass in input image $\mathbf{X}_i$ to $\mathbf{E}(.)$ to obtain $\mathbf{z}$. We project $\mathbf{z}$ unto 2 subspaces using operators: $\{ \mathcal{P}_1; \mathcal{P}_2 \} \in \mathbb{R}^{1024 \times 1024}$. We desire a learned $\{ \mathcal{P}_1; \mathcal{P}_2 \}$ such that $(\mathcal{P}_1 \mathbf{z})^T(\mathcal{P}_2 \mathbf{z}) = 0$. Fig.\ref{fig:latent_space} shows the projection of $\mathbf{z}$ unto orthogonal subspaces. We designate $\mathbf{z}_1$ and $\mathbf{z}_2$ to be the projected latent vectors. Each latent subspace is expected to embed features necessary to disentangle the background of $\mathbf{X}_i$ from its geologic features so we can isolate pixels that belong to background and geologic structure in the pixel space.

\begin{figure}[!htb]
    \centering
    \includegraphics[scale=0.3]{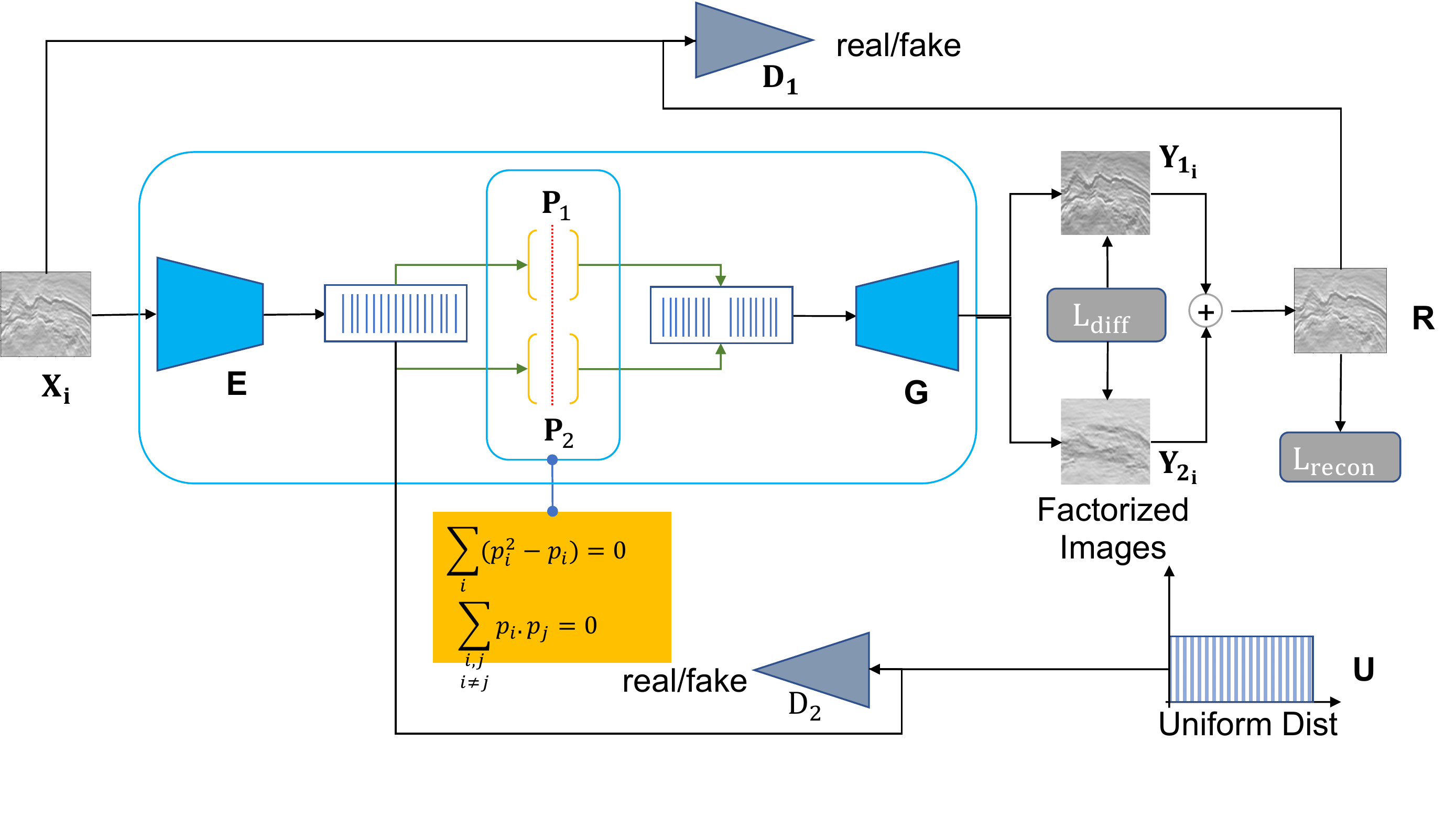}
    \caption{Full model showing discriminators used in adversarial training}
    \label{fig:full_label}
\end{figure}

$\mathbf{G}$ takes in $\mathbf{z}_1, \mathbf{z}_2$ in sequential order and constructs 2 images: $\mathbf{Y}_1 \mbox{ and } \mathbf{Y}_2$. Both constructed images are added together to obtain a reconstruction of the input seismic image, $\mathbf{R} = \mathbf{Y}_1 + \mathbf{Y}_2$. We impose the reconstruction constraint on the model to ensure $\mathbf{Y}_1 \mbox{ and } \mathbf{Y}_2$ does not converge to a weird representation. Invariably, we desire both of them to be 'seismic' plausible qualitatively. Enforcing the reconstruction ensure cycle consistency $\cite{zhu2017unpaired}$.

%% file: adversarial_training.tex
\section{Adversarial Training and Losses}
To enhance our model, we introduce two adversarial training methods on the model. The intuition for adversarial training is obtained from \cite{goodfellow2014generative} and a follow-up work on adversarial auto-encoders by \cite{makhzani2015adversarial}. The first adversarial method is setup to enhance the quality of our reconstructed image, $\mathbf{R}$. Essentially, a discriminator $\mathbf{D}_1$ is associated with $\mathbf{G}$. $\mathbf{G}$ tries to fool $\mathbf{D}_1$ with a reconstruction, $\mathbf{R}_i$ that is similar to an image from the input distribution of $\mathbf{X}_i$ while $\mathbf{D}_1$ tries harder to differentiate $\mathbf{X}_i$ from $\mathbf{R}_i$. The second adversarial method is setup to avoid a mode collapse problem popular with training Generative Adversarial Networks (GANs) by making $\mathbf{z}$ spread out towards a uniform distribution other than a Gaussian distribution. This is done to avoid $\mathbf{z}$ getting stuck on a major mode in the latent space but generalizes on the Gaussian modes in the latent space nicely. Basically, a discriminator $\mathbf{D}_2$ is used to discriminate if the distribution of $\mathbf{z}$ generated by $\mathbf{E}$, is uniform in $[0, 1]$ or it was generated from an arbitrary uniform distribution in $\mathbf{U}_{[0,1]}$ where $\mathbf{U}$ is a uniform vector generator. 

\noindent We define the first adversarial loss as $L_{adv1}$:

\begin{equation}
    \label{l_adv1}
    \begin{aligned}
        \min_{\mathbf{G}} \max_{\mathbf{D}_1} & 
        \mathbb{E}_{\mathbf{x} \sim p_{data}} \left[\log(\mathbf{D}_1(\mathbf{x})) \right] + \\ \mathbb{E}_{\mathbf{z} \sim \mathbf{E}(\mathbf{x})} \left[ \log(1-\mathbf{D}_1(\mathbf{G}(\mathbf{z}))) \right] \\
    \end{aligned}
\end{equation}

\noindent \textit{where $\mathbf{G}$ is the decoder/generator, and $\mathbf{D}_1$ is the first discriminator associated with $\mathbf{G}$.} 
Equation (\ref{l_adv1}) is the adversarial loss between the decoder and the discriminator.

\noindent The second adversarial loss is defined as $L_{adv2}$:
\begin{equation}
\label{l_adv2}
    \begin{aligned}
        \min_{\mathbf{E}} \max_{\mathbf{D}_2}\quad & 
        \mathbb{E}_{\mathbf{z} \sim \mathbf{E}(\mathbf{x})} \left[\log(\mathbf{D}_2(\mathbf{z})) \right] + \\ \mathbb{E}_{\mathbf{u} \sim \mathbf{U}[0,1]} \left[ \log(1-\mathbf{D}_2(\mathbf{u})) \right] \\
    \end{aligned}
\end{equation} \textit{where $\mathbf{U}_{[0,1]}$ generates uniform vectors on [0,1] and $\mathbf{E}$ is the encoder.}

\noindent $L_{adv2}$ is the adversarial loss of not having a uniform latent space as opposed to a Gaussian one. The reconstruction loss on any $\mathbf{X}_i$ is defined as:
\begin{equation}
    \label{eqn:l_mse}
    L_{rec} = \frac{1}{N} \sum_{i=1}^{N} (\mathbf{X}_i - \mathbf{R}_i)^2
\end{equation}
\noindent \textit{where $N$ is the total number of images.} Equation (\ref{eqn:l_mse}) is the $MSE$ of reconstructing $\mathbf{X}$. 
$L_{diff}$ is an $L_1$ loss imposed on $\mathbf{Y}_1$ and $\mathbf{Y}_2$ to ensure they are as different as possible,         but in a sparse sense. Hence their difference is maximized by negating the minimization of the loss during training.              
\begin{equation}
    L_{diff} = -1 \times \sum_{i=1}^N |\mathbf{Y}_{1_i} - \mathbf{Y}_{2_i}|
    \label{ldiff}
\end{equation}

\noindent Lastly, we define $L_{proj}$ as the projection loss over projection matrices.  A projection matrix is idempotent and we need both matrices to be orthogonal to each other. However, the projection of each matrix need not be orthogonal. though. Both idempotency and orthogonality are formulated as constraints as follows:
\begin{equation}
    L_{proj} = \sum_{i,j, \; \; i \neq j}^2 \mathcal{P}_i^T \mathcal{P}_j + \sum_{i=1}^2 (\mathcal{P}^2_i - \mathcal{P}_i) 
\end{equation}

Then, we train the composite model with these losses over $300$ epochs. After each epoch we back-propagate the losses in the following order: $L_{rec}$ is back-propagated to update $\mathbf{E}, \mathbf{G}, \mathcal{P}_1, \mathcal{P}_2$. $L_{adv1}$ is back-propagated to update model $\mathbf{D}_1, \mathbf{G}$. While $L_{adv2}$  is back-propagated to update models $\mathbf{D}_2$ and $\mathbf{E}$. . Finally, $L_{proj}$ and $L_{diff}$ losses are used to update $\mathcal{P}_1, \mathcal{P}_2$.

%% file: results_and_discussion.tex
\section{Results and Discussion}
\begin{figure}[htb]
   \centering
   \label{salt}
   \includegraphics[scale=0.42]{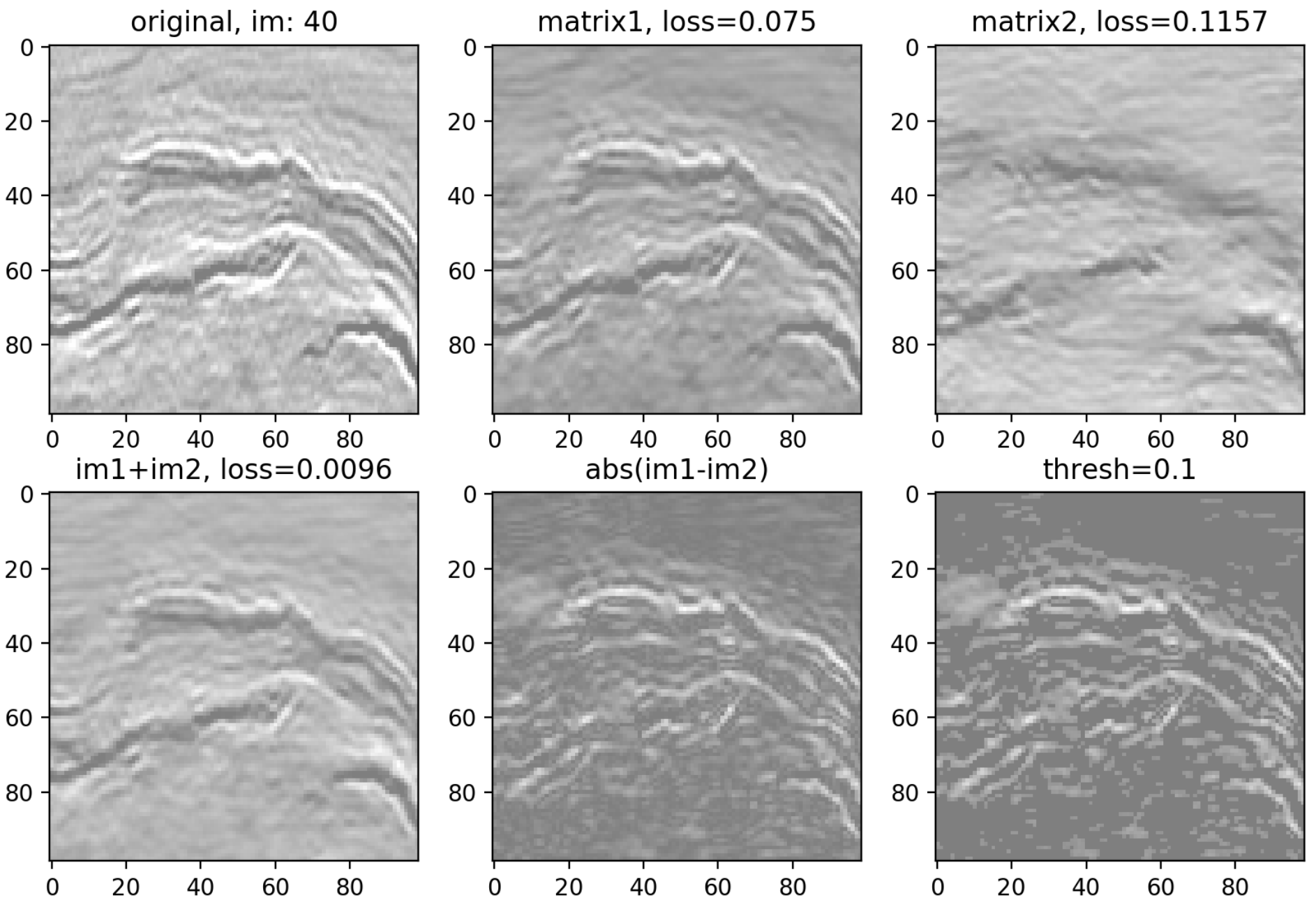}
   \caption{A seismic image taken from a salt-
   dome region. The top left is the input Salt image. The top middle image is image $\mathbf{Y}_1$ from operator $\mathcal{P}_1$. The top right is image $\mathbf{Y}_2$ from operator $\mathcal{P}_2$. The bottom left image is the reconstructed image, $\mathbf{R}$. The bottom middle image is the maximized $l_1$ sparse output from $\mathbf{Y}_1$ and $\mathbf{Y}_2$. The bottom right image is the output annotation showing confidence values for each pixel location.  Notice that the $MSE$ loss between the original image and either $\mathbf{Y}_1$ or $\mathbf{Y}_2$ is less than that of the constructed image. $\mathbf{R}$.}
\end{figure}

Our dataset does not have ground-truth labels. Hence we cannot compute Mean Intersection over Union (MIoU) or similar segmentation metrics. Actually in seismic, to the best of our knowledge, there are no publicly available structurally annotated sections of the F3 block. As such, we make best effort at qualitative comparison. 
When the training is complete, the desired annotated image is the output of minimizing $L_{diff}$ in equation (\ref{ldiff}), which is the same as maximizing the $l_1$ loss between $\mathbf{Y}_{1_i}$ and $\mathbf{Y}_{2_i}$ over 300 epochs. In Fig.~\ref{salt} and Fig.~\ref{horizon}, the annotations (bottom right image) are confidence values, which reflects how much we trust that the corresponding pixel at that location belongs to a geological structure in the original image. 

\begin{figure*}[h!]
\label{comparison}
\centering
  \includegraphics[scale=.70]{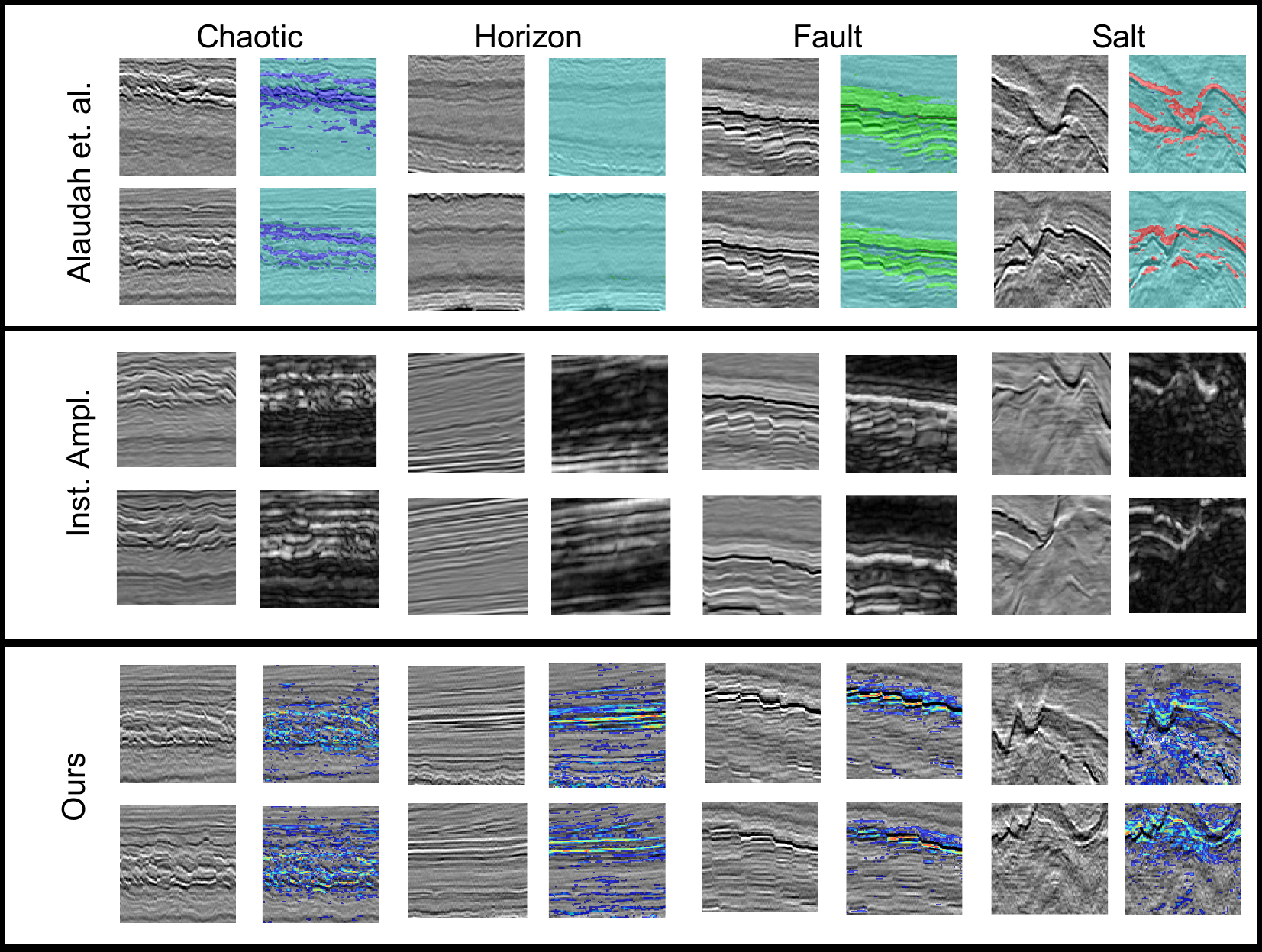}
  \caption{Qualitative comparison of our method with similar annotation frameworks}
\end{figure*}

 \begin{figure}[htb]
       \centering
       \includegraphics[scale=0.30]{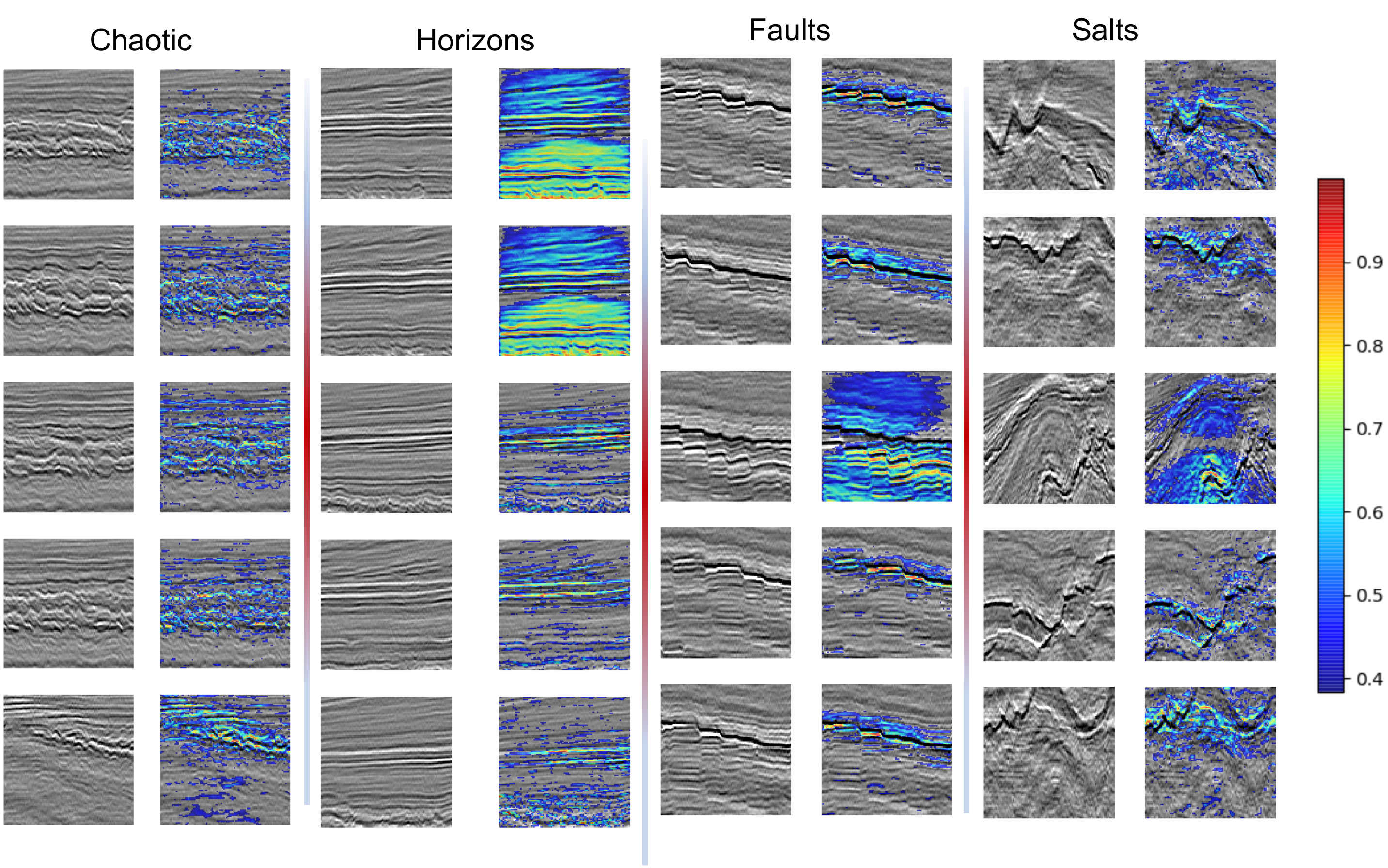}
       \caption{Label Mapping for all 4-Classes}
       \label{fig:label_mapping}
   \end{figure} 

In Fig.5, we compare qualitative performance of our framework against results from \cite{alaudah2018structure} and instantaneous attributes calculated on the whole seismic volume.
The results show that the confidence values associated with our predictions are more precise along the edges of the faulted region, salt, chaotic and horizons. For instance in Alaudah et. al., the whole image is annotated as horizon. In our corresponding horizon image, the lines of the horizon are neatly outlined. In addition, in our salt images, the annotations do not just trace the boundary of the salt as observed in Alaudah et. al., but identifies other non-conformities within the salt body. The instantaneous amplitude images are noisy and the annotations are blurry on the background compared to our annotations.

   \begin{figure}[!htb]
       \centering
       \includegraphics[scale=0.42]{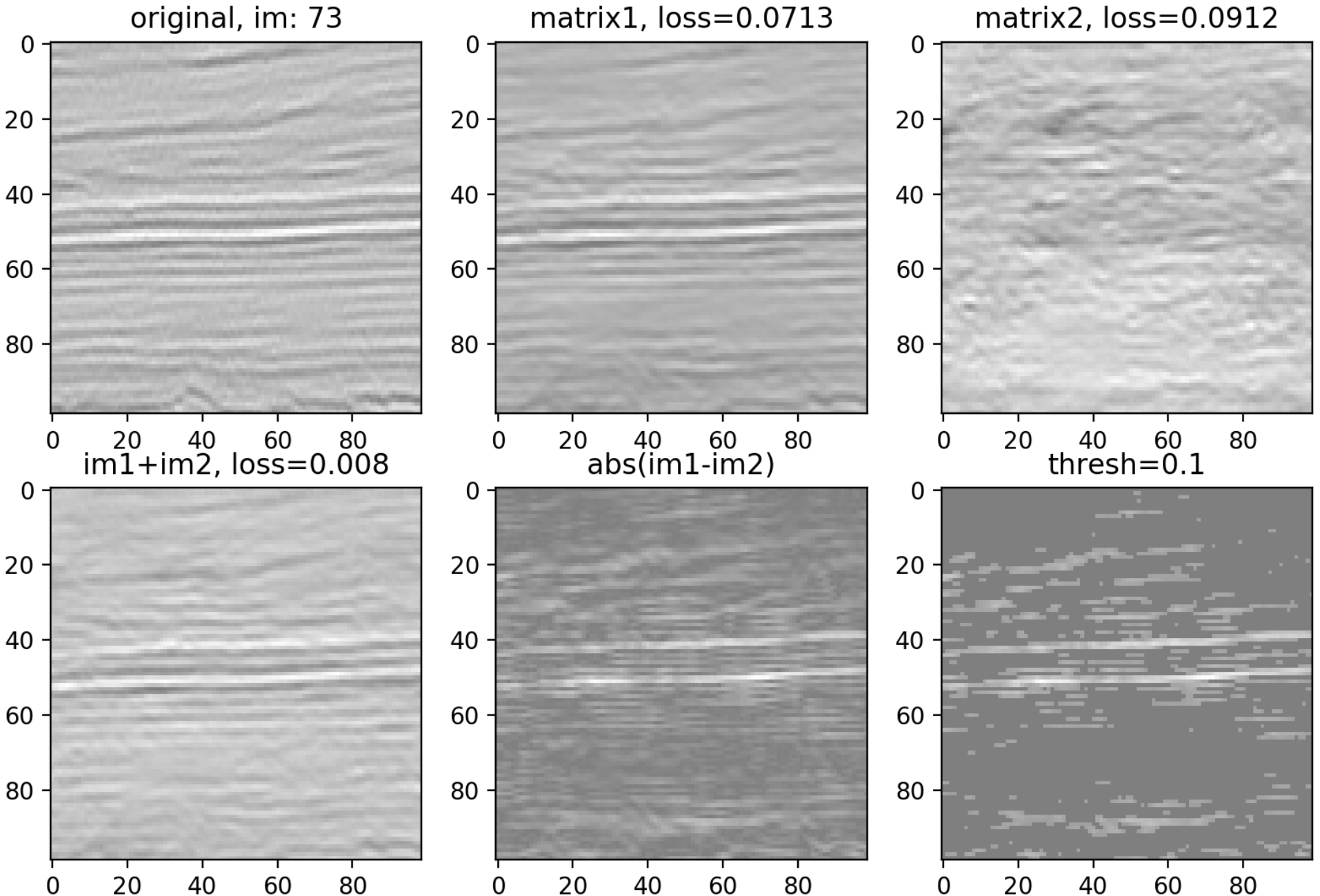}
       \caption{Similar Image to Fig.\ref{salt} except it is shown for Horizon Class}
       \label{horizon}
   \end{figure} 
Fig.~\ref{fig:label_mapping} shows more annotations on various classes in the dataset. Notice in the top 2 images in the Horizon class, there is wide annotation across the image. However, the algorithm designates areas of high confidence with yellow to red makers to indicate the confidence we have in an annotation

\section{Conclusion}
In this paper, we annotated geological elements in seismic images without using strong annotated labels. We rely on defining constraints on a deep adversarial network, to force self-learning of geological regions of interest from the data. The results show that our algorithm annotates the geologically interesting regions of our image. Lastly, we qualitatively compare the results against Alaudah et. al. and a carefully selected seismic attribute (instantaneous attribute) and show how we perform compared to both.

%% file: ms.bbl
\newpage
\begin{thebibliography}{10}

\bibitem{di2019reflector}
Haibin Di and Ghassan AlRegib,
\newblock ``Reflector dip estimates based on seismic waveform curvature/flexure
  analysis,''
\newblock {\em Interpretation}, vol. 7, no. 2, pp. SC1--SC9, 2019.

\bibitem{di2019three}
Haibin Di, Motaz Alfarraj, and Ghassan AlRegib,
\newblock ``Three-dimensional curvature analysis of seismic waveforms and its
  interpretational implications,''
\newblock {\em Geophysical Prospecting}, vol. 67, no. 2, pp. 265--281, 2019.

\bibitem{shafiq2017texture}
Muhammad~Amir Shafiq, Zhen Wang, Ghassan AlRegib, Asjad Amin, and Mohamed
  Deriche,
\newblock ``A texture-based interpretation workflow with application to
  delineating salt domes,''
\newblock {\em Interpretation}, vol. 5, no. 3, pp. SJ1--SJ19, 2017.

\bibitem{shafiq2018role}
Muhammad~Amir Shafiq, Tariq Alshawi, Zhiling Long, and Ghassan AlRegib,
\newblock ``The role of visual saliency in the automation of seismic
  interpretation,''
\newblock {\em Geophysical Prospecting}, vol. 66, no. S1, pp. 132--143, 2018.

\bibitem{alaudah2016generalized}
YK~Alaudah and GI~AlRegib,
\newblock ``A generalized tensor-based coherence attribute,''
\newblock in {\em 78th EAGE Conference and Exhibition 2016}. European
  Association of Geoscientists \& Engineers, 2016, pp. 1--5.

\bibitem{barnes2002investigation}
Arthur~E Barnes and Kenneth~J Laughlin,
\newblock ``Investigation of methods for unsupervised classification of seismic
  data,''
\newblock in {\em SEG Technical Program Expanded Abstracts 2002}, pp.
  2221--2224. Society of Exploration Geophysicists, 2002.

\bibitem{de2007unsupervised}
Marc{\'\i}lio~Castro de~Matos, Paulo~L{\'e}o Osorio, and Paulo~Roberto Johann,
\newblock ``Unsupervised seismic facies analysis using wavelet transform and
  self-organizing maps,''
\newblock {\em Geophysics}, vol. 72, no. 1, pp. P9--P21, 2007.

\bibitem{alaudah2018structure}
Yazeed Alaudah, Motaz Alfarraj, and Ghassan AlRegib,
\newblock ``Structure label prediction using similarity-based retrieval and
  weakly supervised label mapping,''
\newblock {\em Geophysics}, vol. 84, no. 1, pp. V67--V79, 2018.

\bibitem{shafiq2018towards}
Muhammad~A Shafiq, Mohit Prabhushankar, Haibin Di, and Ghassan AlRegib,
\newblock ``Towards understanding common features between natural and seismic
  images,''
\newblock in {\em SEG Technical Program Expanded Abstracts 2018}, pp.
  2076--2080. Society of Exploration Geophysicists, 2018.

\bibitem{dubrovina2019composite}
Anastasia Dubrovina, Fei Xia, Panos Achlioptas, Mira Shalah, Rapha{\"e}l
  Groscot, and Leonidas~J Guibas,
\newblock ``Composite shape modeling via latent space factorization,''
\newblock in {\em Proceedings of the IEEE International Conference on Computer
  Vision}, 2019, pp. 8140--8149.

\bibitem{starck2002curvelet}
Jean-Luc Starck, Emmanuel~J Cand{\`e}s, and David~L Donoho,
\newblock ``The curvelet transform for image denoising,''
\newblock {\em IEEE Transactions on image processing}, vol. 11, no. 6, pp.
  670--684, 2002.

\bibitem{alfarraj2016content}
Motaz Alfarraj, Yazeed Alaudah, and Ghassan AlRegib,
\newblock ``Content-adaptive non-parametric texture similarity measure,''
\newblock in {\em 2016 IEEE 18th International Workshop on Multimedia Signal
  Processing (MMSP)}. IEEE, 2016, pp. 1--6.

\bibitem{long2015characterization}
Zhiling Long*, Yazeed Alaudah, Muhammad~Ali Qureshi, Motaz~Al Farraj, Zhen
  Wang, Asjad Amin, Mohamed Deriche, and Ghassan AlRegib,
\newblock ``Characterization of migrated seismic volumes using texture
  attributes: a comparative study,''
\newblock in {\em SEG Technical Program Expanded Abstracts 2015}, pp.
  1744--1748. Society of Exploration Geophysicists, 2015.

\bibitem{zhu2017unpaired}
Jun-Yan Zhu, Taesung Park, Phillip Isola, and Alexei~A Efros,
\newblock ``Unpaired image-to-image translation using cycle-consistent
  adversarial networks,''
\newblock in {\em Proceedings of the IEEE international conference on computer
  vision}, 2017, pp. 2223--2232.

\bibitem{goodfellow2014generative}
Ian Goodfellow, Jean Pouget-Abadie, Mehdi Mirza, Bing Xu, David Warde-Farley,
  Sherjil Ozair, Aaron Courville, and Yoshua Bengio,
\newblock ``Generative adversarial nets,''
\newblock in {\em Advances in neural information processing systems}, 2014, pp.
  2672--2680.

\bibitem{makhzani2015adversarial}
Alireza Makhzani, Jonathon Shlens, Navdeep Jaitly, Ian Goodfellow, and Brendan
  Frey,
\newblock ``Adversarial autoencoders,''
\newblock {\em arXiv preprint arXiv:1511.05644}, 2015.

\end{thebibliography}
